\documentstyle[12pt]{article}
\font\twlmsb =msbm10 scaled \magstep1
\font\egtmsb =msbm8
\font\sevmsb =msbm7
\newfam\msbfam
\textfont\msbfam\twlmsb
\scriptfont\msbfam\egtmsb
\scriptscriptfont\msbfam\sevmsb
\def\thebibliography#1{\bigskip\section*{\centering
References\\}\bigskip\list
{\arabic{enumi}.}{\settowidth\labelwidth{#1}\leftmargin\labelwidth
\advance\leftmargin\labelsep\usecounter{enumi}}
\def\newblock{\hskip .11em plus .33em minus .07em}
\sloppy\clubpenalty4000\widowpenalty4000 \sfcode`\.=1000\relax}

\def\op#1{\mathop{\fam0 #1}\limits}

\newcommand{\Id}{{\rm Id\,}}
\def\Ker{{\rm Ker\,}}
\newcommand{\ben}{\begin{eqnarray}}
\newcommand{\een}{\end{eqnarray}}
\newcommand{\be}{\begin{eqnarray*}}
\newcommand{\ee}{\end{eqnarray*}}
\newcommand{\bea}{\begin{eqalph}}
\newcommand{\eea}{\end{eqalph}}
\newcommand{\cL}{{\cal L}}
\newcommand{\cE}{{\cal E}}
\newcommand{\cH}{{\cal H}}
\newcommand{\cF}{{\cal F}}
\newcommand{\cD}{{\cal D}}
\newcommand{\al}{\alpha}
\newcommand{\bt}{\beta}

\newcommand{\la}{\lambda}
\newcommand{\La}{\Lambda}
\newcommand{\om}{\omega}
\newcommand{\Om}{\Omega}
\newcommand{\m}{\mu}
\newcommand{\g}{\gamma}
\newcommand{\G}{\Gamma}
\newcommand{\ve}{\varepsilon}
\newcommand{\e}{\epsilon}
\newcommand{\th}{\theta}

\newcommand{\si}{\sigma}
\newcommand{\Si}{\Sigma}
\newcommand{\p}{\pi}
\newcommand{\w}{\wedge}

\newcommand{\wt}{\widetilde}
\newcommand{\wh}{\widehat}
\newcommand{\ol}{\overline}
\newcommand{\dr}{\partial}

\newcounter{eqalph}
\newcounter{equationa}

\newenvironment{eqalph}{\stepcounter{equation}
\setcounter{equationa}{\value{equation}}
\setcounter{equation}{0}

\begin{eqnarray}}{\end{eqnarray}
\setcounter{equation}{\value{equationa}}}
\newenvironment{remark}{{\bf Remark.}}{$\Box$\medskip}

\begin{document}
\hbox{}

\centerline{\bf\large HAMILTONIAN FIELD SYSTEMS ON COMPOSITE MANIFOLDS.}
\bigskip

\centerline{\bf Gennadi A Sardanashvily}
\medskip

\centerline{Department of Theoretical Physics, Physics Faculty,}

\centerline{Moscow State University, 117234 Moscow, Russia}

\centerline{E-mail: sard@grav.phys.msu.su}
\bigskip

The multimomentum Hamiltonian formalism is applied to field
systems represented by sections of composite manifolds $Y\to\Si\to X$
where sections of $\Si\to X$ are parameter fields, e.g.,
Higgs fields and gravitational fields. Their values play the role
of coordinate parameters, besides the world coordinates.

\section{Introduction}

We follow the generally accepted geometric description of classical
fields by sections of fibred manifolds $Y\to X.$

\begin{remark}
A fibred manifold
\[\pi:Y\to X\]
is provided with fibred coordinates $(x^\la, y^i)$ where $x^\la$ are
coordinates of the base $X$.
A locally trivial fibred manifold is termed the bundle.
We denote by $VY$ and $V^*Y$ the vertical tangent bundle and the
vertical cotangent bundle of $Y$ respectively.
For the sake of simplicity, the pullbacks
$Y\op\times_XTX$ and $Y\op\times_XT^*X$
are denoted by $TX$ and $T^*X$ respectively.
\end{remark}

The present article is devoted to field systems on
a composite manifold
\begin{equation}
Y\to\Si\to X\label{I1}
\end{equation}
where $Y\to\Si$ is a bundle denoted by $Y_\Si$ and $\Si\to X$ is a
fibred manifold. In gauge theory, composite manifolds
\[P\to P/K\to X\]
(where $P$ is a principal bundle whose structure group is reducible
to its closed subgroup $K$) describe spontaneous symmetry breaking
\cite{2sar}. Global sections of $P/K\to X$ are treated the Higgs fields.

Application of composite manifolds to field theory is
founded on the following speculations. Given
a global section $h$ of $\Sigma$, the restriction $ Y_h$
of $Y_\Si$ to $h(X)$ is a fibred submanifold
of $Y\to X$. There is the 1:1 correspondence between
the global sections $s_h$ of $Y_h$ and the global sections of
the composite manifold (\ref{I1}) which cover $h$.
Therefore, one can say that sections $s_h$ of $Y_h$
describe fields in the presence of a background parameter
field $h$, whereas sections
of the composite manifold $Y$ describe all pairs $(s_h,h)$.
It is important when the bundles $Y_h$ and $Y_{h\neq h'}$ fail to be
equivalent in a sense. The configuration space of these pairs is the
first order jet manifold $J^1Y$ of the composite manifold $Y$ and
their phase space is the Legendre bundle $\Pi$ over $Y$.

In particular, we aim to apply the composite manifold machinery
to gravitation theory which is the gauge
theory with spontaneous breaking of world symmetries
\cite{tsar,3sar,sard10}.

Let $LX$ be the principal bundle of linear frames in tangent spaces to
a world manifold $X^4$. In gravitation theory, its structure group
\[GL_4=GL^+(4,{\bf R})\]
is reduced to the connected Lorentz group $L=SO(3,1)$.
It means that there exists a reduced subbundle $L^hX$ of $LX$ whose
structure group is $L$. In accordance with the well-known theorem, there is
the 1:1 correspondence between the reduced $L$-principal subbundles $L^hX$ of
$LX$ and the global sections $h$ of the quotient bundle
\begin{equation}
\Si:=LX/L\to X^4.\label{5.15}
\end{equation}
These sections are exactly the tetrad gravitational fields.

The underlying physical reason of spontaneous symmetry breaking in
gravitation theory is Dirac fermion matter
possesing only exact Lorentz symmetries.
The crusial point consists in the fact that,
Dirac fermion field must be regarded only in a pair with a certain
tetrad field $h$. There is the 1:1 correspondence between
these pairs and the sections of the composite bundle
\begin{equation}
S\to\Si\to X^4\label{L1}
\end{equation}
where $S\to\Si$ is a spinor bundle associated with the $SL(2,{\bf
C})$-lift of the $L$-principal
bundle $LX\to\Si$ \cite{3sar,sard10}.
The goal consists in modification of the standard gravitational
equations for sections of the composite manifold (\ref{L1}).

Dynamics of fields represented by sections of a fibred
manifold $Y\to X$ is phrased in terms of jet manifolds
\cite{gia,got,kup,sard,lsar}. In field theory, we
can restrict our consideration to the first order Lagrangian formalism
where the jet manifold $J^1Y$ plays the role of a finite-dimensional
configuration space of fields.

\begin{remark}
The $k$-order jet manifold $J^kY$ of a fibred
manifold $Y$ comprises the equivalence classes
$j^k_xs$, $x\in X$, of sections $s$ of $Y$ identified by the $(k+1)$
terms of their Taylor series at $x$.
The first order jet manifold $J^1Y$ of
$Y$ is both the fibred manifold $J^1Y\to X$
and the affine bundle $J^1Y\to Y $  modelled on the vector
bundle $T^*X\op\otimes_Y VY$.
It is endowed with the adapted coordinates $(x^\la,y^i,y^i_\la)$:
\[{y'}^i_\la=(\frac{\dr {y'}^i}{\dr y^j}y_\m^j +
\frac{\dr{y'}^i}{\dr x^\m})\frac{\dr x^\m}{\dr{x'}^\la}.\]
We identify $J^1Y$ to its image under the  canonical bundle monomorphism
\ben &&\la:J^1Y\op\to_YT^*X \op\otimes_Y TY,\nonumber\\
&&\la=dx^\la\otimes(\dr_\la+y^i_\la \dr_i).\label{18}\een
Given a fibred morphism of $\Phi: Y\to Y'$
over a diffeomorphism of $X$, its jet prolongation
$J^1\Phi:J^1Y\to J^1Y'$ reads
\[ {y'}^i_\mu\circ
J^1\Phi=(\dr_\la\Phi^i+\dr_j\Phi^iy^j_\la)\frac{\dr x^\la}{\dr {x'}^\mu}.\]
A section $\ol s$ of the fibred jet manifold $J^1Y\to X$ is called
holonomic if it is the jet prolongation $\ol s=J^1s$ of a section $s$ of $Y$.
\end{remark}

A Lagrangian density on the configuration space $J^1Y$ is defined to be
a morphism
\be &&L:J^1Y\to\op\w^nT^*X, \qquad n=\dim X,\\
&&L=\cL\om, \qquad \om=dx^1\w...\w dx^n.\ee
Note that since the jet bundle $J^1Y\to Y$ is affine, every polynomial
Lagrangian density of field theory factors
\begin{equation}
L:J^1Y\to T^*X\op\otimes_YVY\to\op\w^nT^*X.\label{523}
\end{equation}

Dynamics of field systems utilizes the language of differential geometry
because of the 1:1 correspondence between the connections on $Y\to X$
and global sections
\begin{equation}
\G =dx^\la\otimes(\dr_\la+\G^i_\la\dr_i)\label{M5}
\end{equation}
of the affine jet bundle $J^1Y\to Y$. These global sections form the
affine space modelled on the linear space of soldering forms on $Y$.
Every connection $\G$ on $Y\to X$ yields the first order differential
operator
\be && D_\G:J^1Y\op\to_YT^*X\op\otimes_YVY,\\
&&D_\G=(y^i_\la-\G^i_\la)dx^\la\otimes\dr_i,\ee
on $Y$ which is called the covariant differential relative to the
connection $\G$.

The feature of the dynamics of field systems on composite manifolds
consists in the following.

Let $Y$ be a composite manifold (\ref{I1})
provided with the fibred coordinates $(x^\la,\si^m,y^i) $ where
$(x^\la,\si^m)$ are fibred coordinates of $\Si$. Every connection
\begin{equation}
A_\Si=dx^\la\otimes(\dr_\la+\wt A^i_\la\dr_i)
+d\si^m\otimes(\dr_m+A^i_m\dr_i) \label{S11}
\end{equation}
on $Y\to\Si$ yields the first order differential operator
\be &&\wt D:J^1Y\to T^*X\op\otimes_Y VY_\Si,\\
&&\wt D=dx^\la\otimes(y^i_\la-\wt A^i_\la -A^i_m\si^m_\la)\dr_i,\ee
on $Y$. Let $h$ be a global section
of $\Si$ and $Y_h$ the restriction of the bundle $Y_\Si$ to $h(X)$. The
restriction of $\wt D$ to $J^1Y_h\subset J^1Y$
comes to the familiar covariant differential relative to a certain
connection $A_h$ on $Y_h$. Thus, it is $\wt D$ that
we may utilize in order to construct a Lagrangian density
\begin{equation}
L:J^1Y\op\to^{\wt D}T^*X\op\otimes_YVY_\Si\to\op\w^nT^*X \label{229}
\end{equation}
for sections of a composite manifold. It should be noted that such a
Lagrangian density is never regular because of the constraint conditions
\[A^i_m\dr^\mu_i\cL=\dr^\mu_m\cL.\]

If a Lagrangian density is degenerate, the corresponding
Euler-Lagrange equations are underdetermined.
To describe constraint field systems, one
can utilize the Hamiltonian formalism in fibred manifolds where
canonical momenta correspond to derivatives of fields with respect
to all world coordinates, not only the temporal one
\cite{car,gun,6sar,sard,lsar}. In the framework of this approach, the phase
space of fields is the Legendre bundle
\begin{equation}
\Pi=\op\w^n T^*X\op\otimes_Y TX\op\otimes_Y V^*Y \label{00}
\end{equation}
over $Y$. It is provided with the fibred coordinates $(x^\la,y^i,p^\la_i)$.
Note that
every Lagrangian density  $L$ on $J^1Y$ determines the Legendre morphism
\be &&\wh L:J^1Y\to \Pi,\\
&&(x^\m,y^i,p^\m_i)\circ\wh L=(x^\m,y^i,\dr^\mu_i\cL).\ee

The Legendre bundle (\ref{00}) carries the multisymplectic form
\begin{equation}
\Om =dp^\la_i\w dy^i\w\om\otimes\dr_\la.\label{406}
\end{equation}
We say that a connection
$\g$ on the fibred Legendre manifold $\Pi\to X$ is a Hamiltonian
connection if the  form  $\g\rfloor\Om$ is closed.
Then, a Hamiltonian form $H$ on $\Pi$ is defined to be an
exterior form such that
\begin{equation}
dH=\g\rfloor\Om \label{013}
\end{equation}
for some Hamiltonian connection $\g$. The key point consists in the
fact that every Hamiltonian form admits splitting
\begin{equation}
H =p^\la_idy^i\w\om_\la -p^\la_i\G^i_\la\om
-\wt{\cH}_\G\om=p^\la_idy^i\w\om_\la-\cH\om
\qquad \om_\la=\dr_\la\rfloor\om.\label{017}
\end{equation}
where $\G$ is a connection on the fibred manifold $Y$ and
$\wt{\cH}_\G\om$ is a horizontal density on $\Pi\to X$.
Given the  Hamiltonian form (\ref{017}), the equality
(\ref{013}) comes to the Hamilton equations
\bea &&\dr_\la r^i=\dr^i_\la\cH, \label{3.11a}\\
&&\dr_\la r^\la_i=-\dr_i\cH \label{3.11b}\eea
for sections $r$ of the fibred Legendre manifold $\Pi\to X$.

If a Lagrangian density $L$ is regular, there exists the unique Hamiltonian
form $H$ such that the first order Euler-Lagrange equations and the
Hamilton equations are equivalent, otherwise in general case.
One must consider a
family of different  Hamiltonian forms $H$ associated with the same
degenerate Lagrangian
density $L$ in order to exaust solutions of the Euler-Lagrange equations.
Lagrangian densities of field models are almost always quadratic and
affine in derivative coordinates $y^i_\mu$. In this case,
given an associated Hamiltonian form $H$, every solution of the
corresponding Hamilton equations which
lives on $\wh L(J^1Y)\subset\Pi$ yields
a solution of the Euler-Lagrange equations. Conversely,
for any solution of the Euler-Lagrange equations, there
exists the corresponding solution of the Hamilton equations for some
associated Hamiltonian form. Obviouslyit lives on  $\wh L(J^1Y)$ which
makes the sense of the Lagrangian constraint space.

The feature of Hamiltonian systems on composite manifolds (\ref{I1})
lies in the facts that: (i) every connection
$A_\Si$ on $Y\to\Si$ yields splitting
\[\om\otimes\dr_\la\otimes
[p^\la_i(dy^i-A^i_md\si^m)+(p^\la_m+A^i_mp^\la_i)d\si^m] \]
of the Legendre bundle $\Pi$ over a composite manifold $Y$ and (ii) the
Lagrangian constraint space is
\begin{equation}
p^\la_m+A^i_mp^\la_i=0. \label{502}
\end{equation}
Moreover, if $h$ is a global section of $\Si\to X$, the submanifold
$\Pi_h$ of $\Pi$ given by the coordinate relations
\[\si^m=h^m(x), \qquad p^\la_m+A^i_mp^\la_i=0\]
is isomorphic to the Legendre bundle over the restriction $Y_h$ of
$Y_\Si$ to $h(X)$. The Legendre bundle $\Pi_h$ is the phase space of
fields in the presence of the background parameter field $h$.

In the Hamiltonian gravitation theory, the constraint condition (\ref{502})
takes the form
\begin{equation}
p^{c\la}_\mu+\frac18\eta^{cb}\si^a_\mu(y^B[\g_a,\g_b]^A{}_B
p^\la_A+p^{A\la}_+[\g_a,\g_b]^{+B}{}_Ay^+_B)=0 \label{M2}
\end{equation}
where $(\si^\mu_c,y^A)$ are tetrad and spinor coordinates of the
composite spinor bundle (\ref{L1}), $p^{c\la}_\mu$ and
$p^\la_A$ are the corresponding momenta and $\eta$ denotes the
Minkowski metric. The condition (\ref{M2}) replaces
the standard gravitational constraints
\begin{equation}
p^{c\la}_\mu=0. \label{M1}
\end{equation}
The crusial point is that, when restricted
to the constraint space (\ref{M1}), the Hamilton equations
come to the familiar gravitational equations, otherwise on the constraint space
(\ref{M2}).

\section{Composite  manifolds}

The composite manifold is defined to be composition of surjective submersions
\begin{equation}
\pi_{\Si X}\circ\pi_{Y\Si}:Y\to \Si\to X. \label{1.34}
\end{equation}
It is provided with the particular class of coordinate atlases
$( x^\la ,\si^m,y^i)$
where $(x^\m,\si^m)$ are fibred  coordinates  of
$\Si$ and $y^i$ are bundle coordinates of $Y_\Si$.
We further propose that $\Si$ has a global section.

Recall the following assertions \cite{sard,sau}:

(i) Let $Y$ be the composite manifold (\ref{1.34}). Given a section $h$ of
$\Si$ and a section $s_\Si$ of $Y_\Si$, their
composition  $s_\Si\circ h$ obviously
is a section of the composite  manifold $Y$. Conversely, if the bundle
$Y_\Si$ has a global section, every
global section $s$ of the fibred manifold $Y\to X$ is  represented by some
composition $s_\Si\circ h$ where $h=\pi_{Y\Si}\circ s$ and $s_\Si$ is an
extension of the local section $h(X)\to s(X)$ of the bundle
$Y_\Si$ over the closed imbedded submanifold $h(X)\subset\Si$.

(ii) Given a global section $h$ of $\Si$, the
restriction $Y_h=h^*Y_\Si$
of the bundle $Y_\Si$ to $h(X)$ is a fibred imbedded submanifold of $Y$.

(iii) There is the 1:1 correspondence between the
sections $s_h$ of $Y_h$ and the sections $s$ of
the composite manifold $Y$ which cover $h$.

Given fibred coordinates $(x^\la, \si^m, y^i)$ of the composite manifold $Y$,
the jet manifolds $J^1\Si$,
$J^1Y_\Si$ and $J^1Y$ are coordinatized respectively by
\[ ( x^\la ,\si^m, \si^m_\la),\qquad
( x^\la ,\si^m, y^i, \wt y^i_\la, y^i_m),\qquad
( x^\la ,\si^m, y^i, \si^m_\la ,y^i_\la).\]

(iv) There exists  the  canonical surjection
\ben &&\rho : J^1\Si\op\times_\Si J^1Y_\Si\to J^1Y, \label{1.38}\\
&&y^i_\la\circ\rho=y^i_m{\si}^m_{\la} +\wt y^i_{\la},\nonumber\een
where $s_\Si$ and $h$ are sections of $Y_\Si$ and $\Si$ respectively.

The following assertions are concerned with connections on composite
manifolds.

Let $A_\Si$ be the connection (\ref{S11}) on the bundle $Y_\Si$ and
$\G$ the connection (\ref{M5}) on the fibred manifold $\Si$.
Building on the morphism (\ref{1.38}), one can construct the composite
connection
\begin{equation}
A=dx^\la\otimes[\dr_\la+\G^m_\la\dr_m +(A^i_m\G^m_\la + \wt A^i_\la)
\dr_i] \label{1.39}
\end{equation}
on the composite manifold $Y$.

Let a global section $h$ of
$\Si$ be an integral section of the connection $\G$ on
$\Si$, that is, $\G\circ h=J^1h$.
Then, the composite connection (\ref{1.39}) on $Y$ is reducible to the
connection
\begin{equation}
A_h=dx^\la\otimes[\dr_\la+(A^i_m\dr_\la h^m
+\wt A^i_\la)\dr_i] \label{1.42}
\end{equation}
on the fibred submanifold $Y_h$ of $Y\to X$.
In particular, every connection $A_\Si$ (\ref{S11}) on $Y_\Si$,
whenever $h$, is reducible to the connection (\ref{1.42}) on $Y_h$.

Every connection (\ref{S11}) on the bundle $Y_\Si$ determines
the horizontal splitting
\begin{equation}
VY=VY_\Si\op\oplus_Y (Y\op\times_\Si V\Si),\label{46}
\end{equation}
\[\dot y^i\dr_i + \dot\si^m\dr_m=
(\dot y^i -A^i_m\dot\si^m)\dr_i + \dot\si^m(\dr_m+A^i_m\dr_i),\]
and the dual horizontal splitting
\begin{equation}
V^*Y=V^*Y_\Si\op\oplus_Y (Y\op\times_\Si V^*\Si),\label{46'}
\end{equation}
\[\dot y_i dy^i + \dot\si_m d\si^m=
\dot y_i(dy^i -A^i_m d\si^m) + (\dot\si_m +A^i_m\dot y_i) d\si^m.\]

Building on the horizontal splitting (\ref{46}), one can constract
the following first order differential operator on the composite
manifold $Y$:
\ben &&\wt D={\rm pr}_1\circ D_A
: J^1Y\to T^*X\op\otimes_Y VY \to T^*X\op\otimes_Y VY_\Si,\nonumber\\
&&\wt D= dx^\la\otimes[y^i_\la-A^i_\la
-A^i_m(\si^m_\la-\G^m_\la)]\dr_i = dx^\la\otimes(y^i_\la-\wt
A^i_\la -A^i_m\si^m_\la)\dr_i, \label{7.10} \een
where $D_A$ is the covariant differential relative to the
composite connection $A$ (\ref{1.39})
which is composition of $A_\Si$ and some
connection $\G$ on $\Si$. We shall call $\wt D$ the vertical covariant
differential. This possesses the following property.

Given a global section $h$ of $\Si$, let $\G$ be a connection on $\Si$
whose integral section is $h$.
It is readily observed that the vertical covariant differential
(\ref{7.10}) restricted to $J^1Y_h\subset J^1Y$
comes to the familiar covariant
differential relative to the connection $A_h$ (\ref{1.42}) on $Y_h$.
Thus, it is the vertical covariant differential (\ref{7.10}) that
we may utilize in order to construct a Lagrangian density (\ref{229})
for sections of a composite manifold.

Now, we consider the composite structure of principal bundles. Let
\[\pi_P: P\to X\]
be a principal bundle with a structure
Lie group $G$ and $K$ its closed subgroup. We have the composite manifold
\begin{equation}
\pi_{\Si X}\circ\pi_{P\Si}:P\to P/K\to X \label{7.16}
\end{equation}
where \[P_\Si:=P\to P/K\]
is a principal bundle with the structure group $K$ and
\[\Si=P/K=(P\times G/K)/G\]
is the $P$-associated bundle. Note that (\ref{7.16}) fails to be a
principal bundle.

Let the structure group $G$ be reducible to its closed subgroup $K$. By the
well-known theorem, there is the 1:1 correspondence
\[\pi_{P\Si}(P_h)=(h\circ\pi_P)(P_h)\]
between global sections $h$ of the
bundle $P/K\to X$ and the reduced $K$-principal
subbundles $P_h$ of $P$ which consist with
restrictions of the principal bundle $P_\Si$ to $h(X)$.

Recall the following facts.
Every principal connection $A_h$ on a reduced subbundle
$P_h$ gives rise to a principal connection on $P$. Conversely, a
principal connection $A$ on $P$ is reducible to a principal connection
on $P_h$ iff $h$ is an integral section of the connection $A$.
Every principal connection $A_\Si$ on the $K$-principal bundle $P_\Si$,
whenever $h$, induces a principal connection on the reduced subbundle
$P_h$ of $P$.

Given the composite manifold (\ref{7.16}), the
canonical morphism (\ref{1.38}) results in the surjection
\[J^1P_\Si/K\op\times_\Si J^1\Si \to J^1P/K \]
over $J^1\Si$. Let $A_\Si$ be a principal connection on $P_\Si$
and $\G$ a connection on $\Si$. The corresponding composite connection
(\ref{1.39}) on the composite manifold (\ref{7.16})
is equivariant under the canonical
action of $K$ on $P$. If the connection $\G$ has an integral global section
$h$ of $P/K\to X$, the composite connection (\ref{1.39}) is reducible to the
connection (\ref{1.42}) on $P_h$ which consists with the principal connection
on $P_h$ induced by $A_\Si$.

Let us consider the composite manifold
\begin{equation}
Y=(P\times V)/K\to P/K\to X \label{7.19}
\end{equation}
where the bundle
\[Y_\Si:=Y_\Si = (P\times V)/K\to P/K\]
is associated with the $K$-principal bundle
$P_\Si$. Given a reduced subbundle $P_h$ of $P$, the associated bundle
\[Y_h=(P_h\times V)/K\]
is isomorphic to the restriction of $Y_\Si$ to $h(X)$.
The composite manifold (\ref{7.19}) can be provided with
the composite connection (\ref{1.39})
where the connection $A_\Si$ is associated with a
principal connection on the $K$ principal bundle
$P_\Si$ and the connection $\G$ on $P/K$ is associated with
a principal connection on some reduced subbundle $P_h$ of $P$. This
composite connection is reducible to the connection (\ref{1.42}) on the bundle
$Y_h$ which appears to be some principal connection $A_h$
on $P$.

\section{Composite spinor bundles}

This Section is devoted to composite spinor bundles (\ref{L1}) in
gravitation theory.

By $X^4$ is further meant an oriented world manifold which satisfies
the well-known global topological conditions in order that gravitational
fields, space-time structure and spinor structure can exist. To
summarize these conditions, we assume that $X^4$ is not compact and
the linear frame bundle $LX$ over $X^4$ is trivial \cite{3sar}.

We describe Dirac fermion fields as follows. Given
a Minkowski space $M$ with the Minkowski metric $\eta$, let
${\bf C}_{1,3}$ be the complex Clifford algebra generated by elements
of $M$.
A spinor space $V$ is defined to be a
linear space of some minimal left ideal of ${\bf C}_{1,3}$  on
which this algebra acts on the left. We have the representation
\begin{equation}
\g: M\otimes V \to V \label{521}
\end{equation}
of elements of the Minkowski space $M\subset{\bf C}_{1,3}$ by
Dirac's matrices $\g$ on $V$.

Let us consider the transformations preserving the representation (\ref{521}).
These are pairs $(l,l_s)$ of Lorentz transformations $l$ of  the Minkowski
space $M$ and invertible elements $l_s$ of ${\bf C}_{1,3}$ such that
\[\g (lM\otimes l_sV)=l_s\g (M\otimes V).\]
Elements $l_s$  form the Clifford group whose action on $M$
however is not effective. We restrict ourselves to its spinor
subgroup $L_s =SL(2,{\bf C})$
whose generators act on $V$ by the representation
\[I_{ab}=\frac{1}{4}[\g_a,\g_b].\]

Let us consider a bundle of complex Clifford algebras ${\bf C}_{3,1}$
over $X^4$. Its subbundles are both a spinor bundle $S_M\to X^4$ and the
bundle $Y_M\to X^4$ of Minkowski spaces of generating elements of
${\bf C}_{3,1}$.
To describe Dirac fermion fields on a world manifold, one must
require $Y_M$ be isomorphic to the cotangent bundle $T^*X$
of a world manifold $X^4$. It takes place if the structure group of
$LX$ is reducible to the Lorentz group $L$ and $LX$
contains a reduced $L$ subbundle $L^hX$ such that
\[Y_M=(L^hX\times M)/L.\]
In this case, the spinor bundle $S_M$ is associated with the $L_s$-lift
$P_h$ of $L^hX$:
\begin{equation}
S_M=S_h=(P_h\times V)/L_s.\label{510}
\end{equation}

There is the above-mentioned 1:1 correspondence between the reduced
subbubdles $L^hX$ of $LX$ and
the tetrad gravitational fields $h$ identified with global sections
of the bundle $\Si$ (\ref{5.15}).

Given a tetrad field $h$, let $\Psi^h$ be an atlas of
$LX$ such that the corresponding local
sections $z_\xi^h$ of $LX$ take their values into $L^hX$.
With respect to $\Psi^h$ and a
holonomic atlas $\Psi^T=\{\psi_\xi^T\}$ of $LX$, a tetrad field $h$
can be represented by a family of $GL_4$-valued tetrad functions
\ben && h_\xi=\psi^T_\xi\circ z^h_\xi,\nonumber\\
&&dx^\la= h^\la_a(x)h^a. \label{L6}\een

Given a tetrad field $h$, one can define the representation
\begin{equation}
\g_h: T^*X\otimes S_h=(P_h\times (M\otimes V))/L_s\to (P_h\times
\g(M\otimes V))/L_s=S_h \label{L4}
\end{equation}
of cotangent vectors to a world manifold $X^4$ by Dirac's $\g$-matrices
on elements of the spinor bundle $S_h$. With respect to an atlas
$\{z_\xi\}$ of $P_h$ and the associated atlas $\{z^h_\xi\}$ of $LX$,
the morphism (\ref{L4}) reads
\[\g_h(h^a\otimes y^Av_A(x))=\g^{aA}{}_By^Bv_A(x)\]
where $\{v_A(x)\}$ are the associated fibre bases
for $S_h$. As a shorthand, one can write
\[\wh dx^\la=\g_h(dx^\la)=h^\la_a(x)\g^a.\]

We shall say that, given the representation (\ref{L4}), sections of
the spinor bundle $S_h$ describe Dirac fermion fields in the presence of
the tetrad gravitational field $h$. Indeed,
let $A_h$ be a principal connection on $S_h$ and
\be &&D: J^1S_h\op\to_{S_h}T^*X\op\otimes_{S_h}VS_h,\\
&&D=(y^A_\la-A^{ab}{}_\la (x)I_{ab}{}^A{}_By^B)dx^\la\otimes\dr_A,\ee
the corresponding covariant differential. Given the
representation (\ref{L4}), one can construct the Dirac operator
\begin{equation}
\cD_h=\g_h\circ D: J^1S_h\to T^*X\op\otimes_{S_h}VS_h\to VS_h, \label{I13}
\end{equation}
\[\dot y^A\circ\cD_h=h^\la_a(x)\g^{aA}{}_B(y^B_\la-A^{ab}{}_\la
I_{ab}{}^A{}_By^B).\]
We here use the fact that the vertical tangent bundle $VS_h$ admits the
canonical splitting
\[VS_h=S_h\times S_h,\]
and $\g_h$ in the expression (\ref{I13}) is the pullback
\be &&\g_h: T^*X\op\otimes_{S_h}VS_h\op\to_{S_h}VS_h,\\
&&\g_h(h^a\otimes\dot y^A\dr_A)=\g^{aA}{}_B\dot y^B\dr_B,\ee
over $S_h$ of the bundle morphism (\ref{L4}).

For different tetrad fields $h$ and $h'$, Dirac fermion fields are
described by sections of spinor bundles $S_h$ and $S_{h'}$ associated
with $L_s$-lifts $P_h$ and $P_{h'}$ of different reduced $L$-principal
subbundles
of $LX$. Therefore, the representations $\gamma_h$ and $\gamma_{h'}$
(\ref{L4}) are not equivalent \cite{tsar,3sar}.
It follows that a Dirac fermion field must be regarded only in a pair with
a certain tetrad gravitational field. There is the 1:1 correspondence
between these pairs and sections of the composite spinor bundle (\ref{L1}).

In gravitation theory, we have the composite manifold
\begin{equation}
\pi_{\Si X}\circ\pi_{P\Si}:LX\to\Si\to X^4 \label{L3}
\end{equation}
where $\Si$ is the quotient bundle (\ref{5.15}) and
\[LX_\Si:=LX\to\Si\]
is the L-principal bundle.

Building on the double universal covering of the group $GL_4$, one can
perform the $L_s$-principal lift $P_\Si$ of $LX_\Si$ such that
\[P_\Si/L_s=\Si, \qquad LX_\Si=r(P_\Si).\]
In particular, there is imbedding of $P_h$ onto the restriction of $P_\Si$
to $h(X^4)$.

Let us consider the composite spinor bundle (\ref{L1}) where
\[S_\Si= (P_\Si\times V)/L_s\] is
associated with the $L_s$-principal bundle $P_\Si$. It is readily observed
that, given a global section $h$ of $\Si$, the restriction $S_\Si$ to
$h(X^4)$ is the spinor bundle $S_h$ (\ref{510}) whose sections describe Dirac
fermion fields in the presence of the tetrad field $h$.

Let us provide the principal bundle $LX$ with a holonomic atlas
$\{\psi^T_\xi, U_\xi\}$ and the principal bundles $P_\Si$ and $LX_\Si$
with associated atlases $\{z^s_\e, U_\e\}$ and $\{z_\e=r\circ z^s_\e\}$.
With respect to these atlases, the composite spinor bundle is endowed
with the fibred coordinates $(x^\la,\si_a^\mu, y^A)$ where $(x^\la,
\si_a^\mu)$ are fibred coordinates of the bundle $\Si$ such that
$\si^\mu_a$ are the matrix components of the group element
\[GL_4\ni (\psi^T_\xi\circ z_\e)(\si): {\bf R}^4\to {\bf R}^4,
\qquad \si\in U_\e,\qquad \pi_{\Si X}(\si)\in U_\xi.\]
Given a section $h$ of $\Si$, we have
\be &&z^h_\xi (x)= (z_\e\circ h)(x), \qquad h(x)\in U_\e,\qquad x\in U_\xi,\\
&& (\si^\la_a\circ h)(x)= h^\la_a(x),\ee
where $h^\la_a(x)$ are tetrad functions (\ref{L6}).

The jet manifolds $J^1\Si$, $J^1S_\Si$ and $J^1S$ are coordinatized by
\[(x^\la,\si^\mu_a, \si^\mu_{a\la}),\qquad
(x^\la,\si^\mu_a, y^A,\wt y^A_\la, y^A{}^a_\mu),\qquad
(x^\la,\si^\mu_a, y^A,\si^\mu_{a\la}, y^A_\la).\]
Note that, whenever $h$, the jet manifold
$J^1S_h$ is a fibred submanifold of $J^1S\to X^4$ given by the coordinate
relations
\[\si^\mu_a=h^\mu_a(x), \qquad \si^\mu_{a\la}=\dr_\la h^\mu_a(x).\]

Let us consider the bundle of Minkowski spaces
\[(LX\times M)/L\to\Si\]
associated with the $L$-principal bundle $LX_\Si$. Since $LX_\Si$ is
trivial, it is isomorphic to
the pullback $\Si\op\times_X T^*X$ which we denote by the same symbol
$T^*X$. Building on the morphism (\ref{521}), one can define the bundle
morphism
\begin{equation}
\g_\Si: T^*X\op\otimes_\Si S_\Si= (P_\Si\times (M\otimes V))/L_s
\to (P_\Si\times\g(M\otimes V))/L_s=S_\Si, \label{L7}
\end{equation}
\[\wh dx^\la=\g_\Si (dx^\la) =\si^\la_a\g^a,\]
over $\Si$. When restricted to $h(X^4)\subset \Si$,
the morphism (\ref{L7}) comes to the morphism $\g_h$
(\ref{L4}). Because of the canonical vertical splitting
\[VS_\Si =S_\Si\op\times_\Si S_\Si,\]
the morphism (\ref{L7}) yields the corresponding morphism
\begin{equation}
\g_\Si: T^*X\op\otimes_SVS_\Si\to VS_\Si. \label{L8}
\end{equation}

We use this morphism in order to construct the total Dirac
operator on sections of the composite spinor bundle $S$ (\ref{L1}). We are
based on the following fact.

Let \[\wt A=dx^\la\otimes (\dr_\la +\wt A^B_\la\dr_B) + d\si^\mu_a\otimes
(\dr^a_\mu+A^B{}^a_\mu\dr_B) \]
be a connection on the bundle $S_\Si$. It determines the horizontal
splitting (\ref{46}) of the vertical tangent bundle $VS$ and the
vertical covariant differential (\ref{7.10}). The composition of
morphisms (\ref{L8}) and (\ref{7.10}) is the first order differential
operator
\[\cD=\g_\Si\circ\wt D:J^1S\to T^*X\op\otimes_SVS_\Si\to VS_\Si,\]
\[\dot y^A\circ\cD=\si^\la_a\g^{aA}{}_B(y^B_\la-\wt A^B_\la -
A^B{}^a_\mu\si^\mu_{a\la}),\] on $S$.
One can treat it the total Dirac operator since, whenever a
tetrad field $h$, the restriction of $\cD$ to $J^1S_h\subset J^1S$ comes
to the Dirac operator $\cD_h$ (\ref{I13})
with respect to the connection
\[A_h=dx^\la\otimes[\dr_\la+(\wt A^B_\la+A^B{}^a_\mu\dr_\la h^\mu_a)\dr_B]
\] on $S_h$.

\section{Multimomentum Hamiltonian formalism}

Let $\Pi$ be the Legendre bundle (\ref{00}) over a fibred manifold
$Y\to X$. This is the composite  manifold
\[\pi_{\Pi X}=\pi\circ\pi_{\Pi Y}:\Pi\to Y\to X\]
provided with fibred coordinates $( x^\la ,y^i,p^\la_i)$:
\begin{equation}
{p'}^\la_i = J \frac{\dr y^j}{\dr{y'}^i} \frac{\dr
{x'}^\la}{\dr x^\m}p^\m_j, \qquad J^{-1}=\det (\frac{\dr {x'}^\la}{\dr
x^\m}). \label{2.3}
\end{equation}
By $J^1\Pi$ is meant the first order jet manifold of
$\Pi\to X$. It is coordinatized by
\[( x^\la ,y^i,p^\la_i,y^i_{(\m)},p^\la_{i\m}).\]

\begin{remark}
We call by a momentum morphism any bundle morphism
$\Phi:\Pi\to J^1Y$ over $Y$.
Given a momentum morphism $\Phi$, its composition with the
monomorphism (\ref{18})
is represented by the horizontal pullback-valued 1-form
\begin{equation}
\Phi =dx^\la\otimes(\dr_\la +\Phi^i_\la\dr_i)\label{2.7}
\end{equation}
on $\Pi\to Y$. For instance, let $\G$ be a connection on $Y$. Then, the
composition $\wh\G=\G\circ\pi_{\Pi Y}$
is a momentum morphism. The
corresponding form (\ref{2.7}) on $\Pi$ is the pullback
$\wh\G$ of the form $\G$ (\ref{M5}) on $Y$. Conversely,
every momentum morphism $\Phi$ defines
the associated connection $ \G_\Phi =\Phi\circ\wh 0_\Pi$
on $Y\to X$ where $\wh 0_\Pi$ is the global zero section of $\Pi\to Y$.
Every connection $\G$ on $Y$ gives rise to the connection
\begin{equation}
\wt\G =dx^\la\otimes[\dr_\la +\G^i_\la (y)\dr_i +
(-\dr_j\G^i_\la (y)  p^\m_i-K^\m{}_{\nu\la}(x) p^\nu_j+K^\al{}_{\al\la}(x)
p^\m_j)\dr^j_\m]  \label{404}
\end{equation}
on $\Pi\to X$ where $K$ is a linear symmetric connection  on $T^*X$.
\end{remark}

The Legendre manifold $\Pi$ carries the multimomentum Liouville form
\begin{equation}
\th =-p^\la_idy^i\w\om\otimes\dr_\la \label{2.4}
\end{equation}
and the multisymplectic form $\Om$ (\ref{406}).

The Hamiltonian formalism in fibred manifolds is formulated
intrinsically in terms of Hamiltonian connections which play the
role similar to that of Hamiltonian vector fields in the symplectic geometry
\cite{6sar,sard,lsar}.

We say that a  connection
$\g$ on the fibred Legendre manifold $\Pi\to X$ is a Hamiltonian
connection if the exterior form $\g\rfloor\Om$  is closed.
An exterior $n$-form $H$ on the
Legendre manifold $\Pi$ is called a  Hamiltonian form if
there exists a Hamiltonian connection  satisfying the equation (\ref{013}).

Let $H$ be a Hamiltonian form. For any exterior horizontal density
$\wt H=\wt{\cH}\om$ on $\Pi\to X$, the form $H-\wt H$ is a Hamiltonian form.
Conversely, if $H$ and $H'$ are  Hamiltonian forms,
their difference $H-H'$ is an exterior horizontal density on $\Pi\to X$.
Thus, Hamiltonian  forms constitute an affine space
modelled on a linear space of the exterior horizontal densities on
$\Pi\to X$.

Let $\G$ be a connection on $Y\to X$ and $\wt\G$ its lift
(\ref{404}) onto $\Pi\to X$. We have the equality
\[\wt\G\rfloor\Om =d(\wh\G\rfloor\th).\]
A glance at this equality shows that $\wt\G$ is a Hamiltonian
connection and
\[ H_\G=\wh\G\rfloor\th =p^\la_i dy^i\w\om_\la -p^\la_i\G^i_\la\om\]
is a Hamiltonian form. It follows that every
Hamiltonian form on $\Pi$ can be
given by the expression (\ref{017}) where $\G$ is some
connection on $Y\to X$.
Moreover, a Hamiltonian form has the canonical splitting (\ref{017})
as follows.
Given a  Hamiltonian form $H$, the vertical tangent morphism
$VH$ yields the momentum morphism
\[ \wh H:\Pi\to J^1Y, \qquad y_\la^i\circ\wh H=\dr^i_\la\cH,\]
and the associated connection $\G_H =\wh H\circ\wh 0$
on $Y$. As a consequence, we have the canonical splitting
\begin{equation}
H=H_{\G_H}-\wt H.\label{3.8}
\end{equation}

Note that every momentum morphism $\Phi$ represented by
the pullback-valued form (\ref{2.7}) on $\Pi$ yields the associated
Hamiltonian form
\begin{equation}
H_\Phi=\Phi\rfloor\th =p^\la_idy^i\w\om_\la -p^\la_i\Phi^i_\la\om.\label{414}
\end{equation}

The Hamilton operator $\cE_H$ for a Hamiltonian form $H$
is defined to be the first order differential operator
\begin{equation}
\cE_H=dH-\wh\Om=[(y^i_{(\la)}-\dr^i_\la\cH) dp^\la_i
-(p^\la_{i\la}+\dr_i\cH) dy^i]\w\om, \label{3.9}
\end{equation}
where $\wh\Om$
is the pullback of the multisymplectic form $\Om$ onto $J^1\Pi$.

For any connection $\g$ on $\Pi\to X$, we have
\[\cE_H\circ\g =dH-\g\rfloor\Om.\]
It follows that  $\g$  is a Hamiltonian jet field for a
Hamiltonian form $H$ if and only if it takes its values into
$\Ker\cE_H$, that is, satisfies  the algebraic Hamilton equations
\begin{equation}
\g^i_\la =\dr^i_\la\cH, \qquad \g^\la_{i\la}=-\dr_i\cH. \label{3.10}
\end{equation}

Let a Hamiltonian connection has an integral section $r$ of $\Pi\to X$.
Then, the Hamilton equations (\ref{3.10}) are brought into the first
order differential Hamilton equations (\ref{3.11a}) and (\ref{3.11b}).

\section{Constraint field systems}

This Section is devoted to relations between Lagrangian and Hamiltonian
formalisms on fibred manifolds in case of degenerate Lagrangian densities.

\begin{remark}
The repeated jet manifold
$J^1J^1Y$, by definition, is the first order jet manifold of
$J^1Y\to X$. It is provided with the adapted coordinates
$(x^\la ,y^i,y^i_\la ,y_{(\m)}^i,y^i_{\la\m}).$
Its subbundle $ \wh J^2Y$ with $y^i_{(\la)}= y^i_\la$ is called the
sesquiholonomic jet manifold.
The second order jet manifold $J^2Y$ of $Y$ is the subbundle
of $\wh J^2Y$ with $ y^i_{\la\m}= y^i_{\m\la}.$
\end{remark}

Let $Y\to X$ be a fibred manifold and $ L=\cL\om$ a Lagrangian density
on $J^1Y$. One can construct the exterior form
\begin{equation}
\La_L =[y^i_{(\la)}-y^i_\la)d\pi^\la_i +
(\dr_i-\wh\dr_\la\dr^\la_i)\cL dy^i]\w\om,\label{304}
\end{equation}
\[ \la=dx^\la\otimes\wh\dr_\la,\qquad
\wh\dr_\la =\dr_\la +y^i_{(\la)}\dr_i+y^i_{\m\la}\dr^\m_i,\]
on the repeated jet manifold $J^1J^1Y$.
Its restriction to the second order jet manifold $J^2Y$ of $Y$ reproduces
the familiar variational Euler-Lagrange operator
\begin{equation}
\cE_L= [\dr_i-
(\dr_\la +y^i_\la\dr_i+y^i_{\m\la}\dr^\m_i)\dr^\la_i]\cL dy^i\w\om.\label{305}
\end{equation}
The restriction of the form (\ref{304}) to the sesquiholonomic jet manifold
$\wh J^2Y$ defines the sesquiholonomic extension
$\cE'_L$  of the Euler-Lagrange operator (\ref{305}).
It is given by the expression
(\ref{305}), but with nonsymmetric coordinates $y^i_{\m\la}$.

Let $\ol s$ be a section of the fibred jet manifold $J^1Y\to X$ such that
its first order jet prolongation  $J^1\ol s$ takes its values into
$\Ker\cE'_L$. Then, $\ol s$ satisfies the first order
differential Euler-Lagrange equations
\ben &&\dr_\la\ol s^i=\ol s^i_\la, \nonumber\\
&& \dr_i\cL-(\dr_\la+\ol s^j_\la\dr_j
+\dr_\la\ol s^j_\m\dr^\m_j)\dr^\la_i\cL=0. \label{306}\een
They are equivalent to the second order Euler-Lagrange equations
\begin{equation}
\dr_i\cL-(\dr_\la+\dr_\la s^j\dr_j
+\dr_\la\dr_\mu s^j \dr^\m_j)\dr^\la_i\cL=0.\label{2.29}
\end{equation}
for sections $s$ of $Y$ where $\ol s=J^1s$.

Let us restrict our consideration to semiregular Lagrangian
densities $L$ when the preimage $\wh L^{-1}(q)$ of each point of
$q\in Q$ is the connected submanifold of $J^1Y$.

Given a Lagrangian density $L$, the vertical tangent morphism $VL$ of
$L$ yields the Legendre morphism
\be &&\wh L : J^1Y\to \Pi, \\
&& p^\la_i\circ\wh L =\pi^\la_i.\ee

We say that a  Hamiltonian form
$H$ is associated with a Lagrangian density $L$ if $H$ satisfies the relations
\bea &&\wh L\circ\wh H\mid_Q=\Id_Q, \qquad Q=\wh L( J^1Y) \label{2.30a},\\
&& H=H_{\wh H}+L\circ\wh H. \label{2.30b}\eea
Note that different  Hamiltonian forms can be associated with the same
Lagrangian density.

All  Hamiltonian forms associated
with a semiregular Lagrangian density $L$ consist with each other on the
constraint space $Q$,
and the Hamilton operator $\cE_H$ (\ref{3.9}) satisfies the relation
\[ \La_L=\cE_H\circ J^1\wh L. \]
Let a  section $r$ of $\Pi\to X$
be a solution of the Hamilton equations (\ref{3.11a}) and (\ref{3.11b})
for a Hamiltonian form $H$ associated with a semiregular Lagrangian
density $L$. If $r$ lives on the constraint space $Q$, the section
$\ol s=\wh H\circ r$ of $J^1Y\to X$ satisfies the first
order Euler-Lagrange equations (\ref{306}).
Conversely, given a semiregular Lagrangian density $L$, let
$\ol s$ be a solution of the
first order Euler-Lagrange equations (\ref{306}).
Let $H$ be a Hamiltonian form associated with $L$ so that
\begin{equation}
\wh H\circ \wh L\circ \ol s=\ol s.\label{2.36}
\end{equation}
Then, the section $r=\wh L\circ \ol s$ of $\Pi\to X$ is a solution of the
Hamilton equations (\ref{3.11a}) and (\ref{3.11b}) for $H$.
For sections $\ol s$ and $r$, we have the relations
\[\ol s=J^1s, \qquad  s=\pi_{\Pi Y}\circ r\]
where $s$ is a solution of the second order Euler-Lagrange equations
(\ref{2.29}).

We shall say that a family of Hamiltonian forms $H$
associated with a semiregular Lagrangian density $L$ is
complete if, for each solution $\ol s$ of the first order Euler-Lagrange
equations (\ref{306}), there exists
a solution $r$ of the Hamilton equations (\ref{3.11a}) and (\ref{3.11b}) for
some  Hamiltonian form $H$ from this family so that
\[ r=\wh L\circ\ol s,\qquad  \ol s =\wh H\circ r, \qquad
\ol s= J^1(\pi_{\Pi Y}\circ r). \]
Such a complete family
exists iff, for each solution $\ol s$ of the Euler-Lagrange
equations for $L$, there exists a  Hamiltonian form $H$ from this
family so that the condition (\ref{2.36}) holds.

The most of field models possesses affine and
quadratic Lagrangian densities. Complete
families of Hamiltonian forms associated with such Lagrangian densities
always exist \cite{sard}.

As a test case, let us consider the gauge theory of principal connections.

In the rest of this Section, the manifold $X$ is assumed to be
oriented and provided with a nondegenerate fibre metric $g_{\m\nu}$
in the tangent bundle of $X$. We denote $g=\det(g_{\m\nu}).$

Let $P\to X$ be a principal bundle with a structure Lie group $G$
wich acts on $P$ on the right.
There is the 1:1 correspondence between the principal connections $A$ on
$P$  and the global sections of the bundle $C=J^1P/G$.
It is the affine bundle modelled on the vector bundle
\[\ol C =T^*X \otimes V^GP, \qquad  V^GP=VP/G.\]

Given a bundle atlas $\Psi^P$ of $P$, the bundle $C$
is provided with  the fibred coordinates $(x^\mu,k^m_\mu)$ so that
\[(k^m_\mu\circ A)(x)=A^m_\mu(x)\]
are coefficients of the local connection 1-form of a principal connection
$A$ with respect to the atlas $\Psi^P$.
The first order jet manifold $J^1C$ of the bundle $C$ is
provided with the adapted coordinates $
(x^\mu, k^m_\mu, k^m_{\mu\lambda}).$

There exists the canonical splitting
\begin{equation}
J^1C=C_+\op\oplus_C C_-=(J^2P/G)\op\oplus_C
(\op\w^2 T^*X\op\otimes_C V^GP), \label{N31}
\end{equation}
\[ k^m_{\mu\la}=\frac12(k^m_{\mu\la}+k^m_{\la\mu}+c^m_{nl}k^n_\la k^l_\mu)
+\frac12( k^m_{\mu\la}-k^m_{\la\mu} -c^m_{nl}k^n_\la k^l_\mu),\]
over $C$. There are the corresponding canonical surjections:
\be &&{\cal S}: J^1 C\to C_+, \qquad {\cal S}^m_{\la\mu}=
k^m_{\mu\la}+k^m_{\la\mu} +c^m_{nl}k^n_\la k^l_\mu,\\
&& \cF: J^1 C\to C_-,\qquad
\cF^m_{\la\mu}= k^m_{\mu\la}-k^m_{\la\mu} -c^m_{nl}k^n_\la k^l_\mu.\ee

The Legendre bundle over the bundle $C$ is
\[\Pi=\op\w^n T^*X\otimes TX\op\otimes_C [C\times\ol C]^*.\]
It is coordinatized by $(x^\mu,k^m_\mu,p^{\mu\la}_m)$.

On the configuration space (\ref{N31}),
the conventional Yang-Mills Lagrangian density $L_{YM}$
is given by the expression
\begin{equation}
L_{YM}=\frac{1}{4\ve^2}a^G_{mn}g^{\la\mu}g^{\bt\nu}\cF^m_{\la
\beta}\cF^n_{\mu\nu}\sqrt{\mid g\mid}\,\om \label{5.1}
\end{equation}
where  $a^G$ is a nondegenerate $G$-invariant metric
in the Lie algebra of $G$. It is almost regular and semiregular.
The Legendre morphism
associated with the Lagrangian density (\ref{5.1}) takes the form
\bea &&p^{(\mu\la)}_m\circ\wh L_{YM}=0, \label{5.2a}\\
&&p^{[\mu\la]}_m\circ\wh L_{YM}=\ve^{-2}a^G_{mn}g^{\la\al}g^{\mu\bt}
\cF^n_{\al\bt}\sqrt{\mid g\mid}. \label{5.2b}\eea

Let us consider connections on the bundle $C$ which
take their values into $\Ker\wh L_{YM}$:
\begin{equation}
S:C\to C_+, \qquad
S^m_{\m\la}-S^m_{\la\m}-c^m_{nl}k^n_\la k^l_\m=0. \label{69}
\end{equation}

For all these connections, the Hamiltonian forms
\ben &&H=p^{\mu\la}_mdk^m_\mu\w\om_\la-
p^{\mu\la}_mS_B{}^m_{\mu\la}\om-\wt{\cH}_{YM}\om, \label{5.3}\\
&&\wt{\cH}_{YM}= \frac{\ve^2}{4}a^{mn}_Gg_{\mu\nu}
g_{\la\bt} p^{[\mu\la]}_m p^{[\nu\bt]}_n\mid g\mid ^{-1/2},\nonumber\een
are associated with the Lagrangian density $L_{YM}$ and constitute the
complete family.
Moreover, we can minimize this complete family if we restrict our
consideration to connections (\ref{69}) of the following type.
Given a symmetric linear connection $K$
on the cotangent bundle $T^*X$ of $X$,  every principal connection $B$ on
$P$ is lifted to the connection $S_B$ (\ref{69}) such that
\[S_B\circ B={\cal S}\circ J^1B,\]
\[S_B{}^m_{\mu\la}=\frac{1}{2} [c^m_{nl}k^n_\la
k^l_\mu  +\dr_\mu B^m_\la+\dr_\la B^m_\mu -c^m_{nl}
(k^n_\mu B^l_\la+k^n_\la B^l_\mu)] -K^\bt{}_{\mu\la}(B^m_\bt-k^m_\bt).\]

The corresponding Hamilton equations for sections $r$ of $\Pi\to X$ read
\ben &&\dr_\la p^{\mu\la}_m=-c^n_{lm}k^l_\nu
p^{[\mu\nu]}_n+c^n_{ml}B^l_\nu p^{(\mu\nu)}_n
-K^\mu{}_{\la\nu}p^{(\la\nu)}_m, \label{5.5} \\
&&\dr_\la k^m_\mu+ \dr_\mu k^m_\la=2S_B{}^m_{(\mu\la)}\label{5.6}\een
plus the equation (\ref{5.2b}). The
equations (\ref{5.2b}) and (\ref{5.5}) restricted to the constraint space
(\ref{5.2a}) are the familiar
Yang-Mills equations for $A=\pi_{\Pi C}\circ r.$
Different Hamiltonian forms \ref{5.3}) lead to different
equations (\ref{5.6}) which play the role of the gauge-type condition.

\section{Hamiltonian systems on composite manifolds}

The major feature of Hamiltonian systems on a composite manifold
$Y$ (\ref{1.34}) lies in the following.
The horizontal splitting (\ref{46'}) yields immediately
the corresponding splitting
of the Legendre bundle $\Pi$ over the composite manifold $Y$. As a consequence,
the Hamilton equations (\ref{3.11a}) for sections $h$
of the fibred manifold $\Si$ reduce to the gauge-type conditions
independent of momenta.
Thereby, these sections play the role of parameter fields. Their
momenta meet the constraintconditions (\ref{502}).

Let $Y$ be a composite manifold (\ref{1.34}). The Legendre bundle
$\Pi$ over $Y$ is coordinatized by
\[(x^\la,\si^m,y^i,p^\la_m,p^\la_i).\]
Let $A_\Si$ be a connection (\ref{S11}) on the bundle $Y_\Si$.
With a connection $A_\Si$, the splitting
\begin{equation}
\Pi=\op\w^nT^*X\op\otimes_YTX\op\otimes_Y
[V^*Y_\Si\op\oplus_Y (Y\op\times_\Si V^*\Si)]\label{230}
\end{equation}
of the Legendre bundle $\Pi$ is performed as an immediate consequence
of the splitting (\ref{46'}). We call this the horizontal splitting of
$\Pi$. Given the horizontal splitting (\ref{230}), the Legendre
bundle $\Pi$ can be provided with the coordinates
\begin{equation}
\ol p^\la_i=p^\la_i, \qquad \ol p^\la_m=p^\la_m +A^i_mp^\la_i\label{231}
\end{equation}
which are compatible with this splitting.

Let $h$ be a global section of the fibred manifold $\Sigma$.
It is readily observed that, given the horizontal splitting
(\ref{230}), the submanifold
\begin{equation}
\{\si=h(x), \ol p^\la_m=0\}\label{7.11}
\end{equation}
of the Legendre bundle $\Pi$ over $Y$ is isomorphic to
the Legendre bundle $\Pi_h$ over the restriction  $Y_h$ of $Y_\Si$ to $h(X)$.

Let the composite manifold $Y$ be provided with the composite
connection (\ref{1.39}) determined by connections $A_\Si$ on $Y_\Si$
and $\G$ on $\Si$. Relative to the coordinates (\ref{231}),
every Hamiltonian form on the Legendre bundle $\Pi$
over $Y$ can be given by the expression
\begin{equation}
H=(p^\la_idy^i+p^\la_md\si^m)\w\om_\la-
[\ol p^\la_i\wt A^i_\la +\ol p^\la_m\G^m_\la
+\wt{\cH}(x^\mu, \si^m, y^i, \ol p^\mu_m, \ol p^\mu_i)]\om. \label{7.12}
\end{equation}
The corresponding Hamilton equations are written
\bea &&\dr_\la p^\la_i=-p^\la_j[\dr_i\wt A^j_\la
+\dr_iA^j_m(\G^m_\la +\dr^m_\la\wt{\cH})]-\dr_i\wt{\cH},\label{7.13a} \\
&&\dr_\la y^i=\wt A^i_\la +A^i_m(\G^m_\la
+\dr^m_\la\wt{\cH}) +\dr^i_\la\wt{\cH}, \label{7.13b} \\
&&\dr_\la p^\la_m= -p^\la_i[\dr_m\wt A^i_\la
+\dr_mA^i_n(\G^n_\la +\dr^n_\la\wt{\cH})] -\ol
p^\la_n\dr_m\G^m_\la -\dr_m\wt{\cH}, \label{7.13c} \\
&&\dr_\la\si^m=\G^m_\la +\dr^m_\la\wt{\cH} \label{7.13d}\eea
and plus constraint conditions.

In particular, let the Hamiltonian form
(\ref{7.12}) be associated with a Lagrangian density (\ref{229})
which contains the velocities $\si^m_\mu$ only inside the vertical
covariant differential (\ref{7.10}).
Then, the Hamiltonian density $\wt{\cH}\om$ appears independent of
the momenta $\ol p^\m_m$ and the Lagrangian constraints read
\begin{equation}
\ol p^\m_m=0. \label{232}
\end{equation}
In this case, the Hamilton equation (\ref{7.13d})
comes to the gauge-type condition
\[\dr_\la\si^m=\G^m_\la \]
independent of momenta.

Let us consider now a Hamiltonian system in the presence of a background
parameter field $h(x)$. After
substituting the equation (\ref{7.13d}) into the equations (\ref{7.13a})
- (\ref{7.13b}) and restricting them to the submanifold (\ref{7.11}), we
obtain the equations
\ben &&\dr_\la p^\la_i=-p^\la_j\dr_i[(\wt A\circ h)^j_\la
+A^j_m\dr_\la h^m]-\dr_i\wt{\cH},\nonumber \\
&&\dr_\la y^i=(\wt A\circ h)^i_\la +A^i_m\dr_\la h^m
+\dr^i_\la\wt{\cH}  \label{7.14}\een
for sections of the fibred Legendre manifold $\Pi_h\to X$  of the bundle $Y_h$
endowed with the connection (\ref{1.42}). Equations (\ref{7.14}) are the
Hamilton equations corresponding to the Hamiltonian form
\[H_h=p^\la_idy^i\w\om_\la -[p^\la_i A_h{}^i_\la
+\wt{\cH}(x^\mu, h^m(x), y^i, p^\mu_i, \ol p^\mu_m=0)]\om\]
on $\Pi_h$ which is induced by the Hamiltonian form (\ref{7.12}) on $\Pi$.

\section{Hamiltonian gravitation theory}

At first, let us consider Dirac fermion fields in the
presence of a background tetrad field $h$. Recall that they are represented
by global sections of the spinor bundle $S_h$ (\ref{510}).
Their Lagrangian density
is defined on the configuration space $J^1(S_h\oplus S^*_h)$ provided
with the adapted coordinates
$( x^\m, y^A,y^+_A, y^A_\mu, y^+_{A\mu})$.
It is the affine Lagrangian density
\ben &&L_D=\{\frac{i}2[ y^+_A(\g^0\g^\m)^A{}_B( y^B_\m -A^B{}_{C\m}y^C)
-(y^+_{A\m}-A^+{}^C{}_{A\m}y^+_C)(\g^0\g^\m)^A{}_By^B]\nonumber\\
&&\qquad -my^+_A(\g^0)^A{}_By^B\}h^{-1}\om,\qquad
\g^\m =h^\m_a( x)\g^a, \qquad h=\det (h^\mu_a),\label{5.60}\een
where \[A^A{}_{B\mu}=\frac12A^{ab}{}_\mu (x)I_{ab}{}^A{}_B\]
is a principal connection on the principal spinor bundle $P_h$.

The Legendre bundle $\Pi_h$
over the spinor bundle $S_h\oplus S_h^*$ is coordinatized by
\[( x^\m ,y^A,y^+_A,p^\m_A,p^{\m A}_+).\]
Relative to these coordinates, the Legendre morphism associated
with the Lagrangian density (\ref{5.60}) is written
\ben &&p^\m_A=\pi^\m_A=\frac{i}2y^+_B(\g^0\g^\m)^B{}_Ah^{-1},\nonumber\\
&&p^{\m A}_+=\pi^{\m A}_+=-\frac{i}2(\g^0\g^\m)^A{}_By^Bh^{-1}.\label{5.61}
\een
It defines the constraint subspace of the Legendre bundle $\Pi_h$. Given a
soldering form
\[S=S^A{}_{B\m}(x)y^Bdx^\m\otimes\dr_A\]
on the bundle $S_h$,
let us consider the connection $A+S$ on $S_h$. The corresponding
Hamiltonian form associated with the Lagrangian
density (\ref{5.60}) reads
\ben &&H_S=( p^\m_A dy^A+p^{\m A}_+dy^+_A)\w\om_\m-\cH_S\om, \label{N54}\\
&&\cH_S=p^\m_AA^A{}_{B\m}y^B+y^+_BA^+{}^B{}_{A\m}p^{\m
A}_++my^+_A(\g^0)^A{}_By^Bh^{-1}+\nonumber\\
&&\qquad ( p^\m_A-\pi^\m_A) S^A{}_{B\m} y^B+y^+_BS^+{}^B{}_{A\m}( p^{\m
A}_+-\pi^{\m A}_+).\nonumber\een

The corresponding Hamilton equations for a section $r$ of the fibred
Legendre manifold $\Pi_h\to X$ take the form
\bea &&\dr_\m y^+_A=y^+_B( A^+{}^B{}_{A\m}+S^+{}^B{}_{A\m}),\label{5.62a}\\
&&\dr_\m p^\m_A=-p^\m_BA^B{}_{A\m}-( p^\m_B-\p^\m_B) S^B{}_{
A\m}- [my^+_B(\g^0)^B{}_A
+\frac{i}{2}y^+_BS^+{}^B{}_{C\m}(\g^0\g^\m)^C{}_A]h^{-1}\label{5.62b} \eea
plus the equations for the components $y^A$ and $p^{\m A}_+$.
The equation (\ref{5.62a})
and the similar equation for $y^A$ imply that $y$ is an integral
section of the connection $A+S$ on the spinor bundle $S_h$. It
follows that the Hamiltonian forms (\ref{N54})
constitute the complete family. On the constraint
space (\ref{5.61}), the equation (\ref{5.62b}) comes to the form
\begin{equation}
\dr_\m\pi^\m_A=-\pi^\m_BA^B{}_{A\m}-(
my^+_B(\g^0)^B{}_A+ \frac{i}2y^+_BS^+{}^B{}_{C\m}(\g^0\g^\m)^C{}_A)
h^{-1}.\label{5.63}
\end{equation}
Substituting the equation (\ref{5.62a}) into the equation (\ref{5.63}), we
obtain the familiar Dirac equation in the presence of a tetrad
gravitational field $h$.

We now consider gravity without matter.

In the gauge gravitation theory, dynamic
gravitational variables are pairs of
tetrad gravitational fields $h$ and gauge gravitational potentials $A_h$
identified with principal connections on $P_h$. Following general procedure,
one can describe these  pairs $(h, A_h)$
by sections of the composite bundle
\[C_L:=J^1LX/L\to J^1\Si\to\Si\to X^4.\]
The corresponding configuration space is the jet manifold
$J^1C_L$ of $C_L$. The Legendre bundle
\begin{equation}
\Pi=\op\w^4 T^*X^4\op\otimes_{C_L} TX^4\op\otimes_{C_L} V^*C_L.\label{5.54}
\end{equation}
over $C_L$ plays the role of a phase space of the gauge gravitation theory.

The bundle $C_L$ is endowed with the local fibred coordinates
\[(x^\mu,\si^\la_a,k^{ab}{}_\la=-k^{ba}{}_\la, \si^\la_{a\mu})\]
where $(x^\mu,\si^\la_a, \si^\la_{a\mu})$
are coordinates of the jet bundle $J^1\Si$.
The jet manifold $J^1C_L$ of $C_L$ is provided with the corresponding
adapted coordinates
\[(x^\mu,\si^\mu_a,k^{ab}{}_\la=
-k^{ba}{}_\la, \si^\mu_{a\la}=\si^\mu_{a(\la)},
k^{ab}{}_{\mu\la},\si^\mu_{a\la\nu}).\]

The associated coordinates of the  Legendre manifold (\ref{5.54}) are
\[(x^\mu, \si^\la_a, k^{ab}{}_\la, \si^\la_{a\nu},
p^{a\mu}_\la, p_{ab}{}^{\la\mu}, p_\la^{a\nu\mu})\]
where $(x^\mu, \si^\la_a, p^{a\mu}_\la) $ are
coordinates of the Legendre manifold of the bundle $\Sigma$.

For the sake of simplicity, we here consider
the Hilbert-Einstein Lagrangian density of classical gravity
\begin{equation}
L_{HE}=-\frac{1}{2\kappa}\cF^{ab}{}_{\mu\la}\si^\mu_a
\si^\la_b\si^{-1}\om, \label{5.56}
\end{equation}
\[\cF^{ab}{}_{\mu\la}=k^{ab}{}_{\la\mu}-k^{ab}{}_{\mu\la}+
k^a{}_{c\mu} k^{cb}{}_\la-k^a{}_{c\la} k^{cb}{}_\mu,\]
\[ \si=\det(\si^\mu_a).\]
The corresponding Legendre morphism
$\wh L_{HE}$ is given by the coordinate expressions
\bea &&p_{ab}{}^{[\la\mu]}=\pi_{ab}{}^{[\la\mu]}=\frac{-1}{\kappa\si}
\si^{[\mu}_a\si^{\la]}_b. \label{5.57a}\\
&& p_{ab}{}^{(\la\mu)}=0, \qquad p^{a\mu}_\la=0,\qquad
p^{a\nu\mu}_\la=0, \label{5.57b}\eea

We construct the complete family of Hamiltonian forms
associated with the affine Lagrangian density (\ref{5.56}). Let
$K$ be a world connection associated with a principal connection $B$ on the
linear frame bundle $LX$. To minimize the complete family, we consider the
following connections on the bundle $C_K$:
\be &&\G^\la_{a\mu}=B^b{}_{a\mu}\si^\la_b -K^\la{}_{\nu\mu}\si^\nu_a,\\
&&\G^\la_{a\nu\mu}=\dr_\mu B^d{}_{a\nu}\si^\la_d
-\dr_\mu K^\la{}_{\bt\nu}\si^\bt_a \\
&& \qquad
+B^d{}_{a\mu}(\si^\la_{d\nu}-\G^\la_{d\nu})-K^\la{}_{\bt\mu}(\si^\bt_{a\nu}
-\G^\bt_{a\nu})+K^\bt{}_{\nu\mu}(\si^\la_{a\bt}-\G^\la_{a\bt})\\
&& \qquad +B^d{}_{a\nu}\G^\la_{d\mu}-K^\la{}_{\bt\nu}\G^\bt_{a\mu},\\
&&\G^{ab}{}_{\la\mu}=\frac12 [k^a{}_{c\la}k^{cb}{}_\mu-
k^a{}_{c\mu}k^{cb}{}_\la+\dr_\la B^{ab}{}_\mu+\dr_\mu B^{ab}{}_\la \\
&& \qquad -B^b{}_{c\mu}k^{ac}{}_\la -B^b{}_{c\la}k^{ac}{}_\mu
-B^a{}_{c\mu}k^{cb}{}_\la  -B^a{}_{c\la}k^{cb}{}_\mu]\\
&& \qquad +K^\nu{}_{\la\mu}k^{ab}{}_\nu-K^\nu{}_{(\la\mu)}
B^{ab}{}_\nu -\frac12 R^{ab}{}_{\la\m},\ee
where $R$ is the curvature of the connection $B$.

The complete family of Hamiltonian forms associated
with the Lagrangian density (\ref{5.56}) consists of the forms given by the
coordinate expressions
\be &&H_{HE}=(p_{ab}{}^{\la\mu}dk^{ab}{}_\la+p^{a\mu}_\la d\si^\la_a +
p_\la^{a\nu\mu}d\si^\la_{a\nu})\w\om_\mu-\cH_{HE}\om,\\
&&\cH_{HE}=(p_{ab}{}^{\la\mu} \G^{ab}{}_{\la\mu}+
p^{a\mu}_\la\G^\la_{a\mu}+p_\la^{a\nu\mu}\G^\la_{a\nu\mu})+
\frac12 R^{ab}{}_{\la\mu}(p_{ab}{}^{[\la\mu]}-\pi_{ab}{}^{\la\mu}).\ee
The Hamilton equations corresponding to such a Hamiltonian form read
\bea && \cF^{ab}{}_{\mu\la}=R^{ab}{}_{\mu\la}, \label{5.58a}\\
&&\dr_{\mu}k^{ab}{}_{\la}+\dr_{\la}k^{ab}{}_{\mu}
=2\G^{ab}{}_{(\mu\la)}, \label{5.58b}\\
&& \dr_\mu\si^\la_a=\G^\la_{a\mu}, \label{5.58c}\\
&&\dr_\mu\si^\la_{a\nu}=\G^\la_{a\nu\mu}, \label{5.58d} \\
 &&\dr_\mu p_{ac}{}^{\la\mu}=-\frac{\dr\cH_{HE}}{\dr
k^{ac}{}_\la}, \label{5.58e}\\
&&\dr_\mu p^{a\mu}_\la=-\frac{\dr\cH_{HE}}{\dr\si^\la_a},\label{5.58f}\eea
plus the equations which are reduced to the trivial identities
on the constraint space (\ref{5.57a}). The equations (\ref{5.58a}) -
(\ref{5.58d}) make the sense of gauge-type
conditions. The equation (\ref{5.58d}) has the solution
\[\si^\la_{a\mu} =\dr_\nu\si^\la_a.\]
The gauge-type condition (\ref{5.58b}) has the solution
$k(x)=B.$ It follows that the
forms $H_{HE}$ really constitute the complete family of
Hamiltonian forms associated with the Hilbert-Enstein Lagrangian density
(\ref{5.56}).

On the constraint  space, the equations (\ref{5.58e}) and (\ref{5.58f})
are brought into the form
\bea &&\dr_\mu\pi_{ac}{}^{\la\mu}=2k^b{}_{c\mu}\pi_{ab}{}^{\la\mu}+
\pi_{ac}{}^{\bt\g}\G^\la{}_{\bt\g},\label{5.59a}\\
&&R^{cb}{}_{\bt\mu}\dr^a_\la\pi_{cb}{}^{\bt\mu}=0.\label{5.59b}\eea
The equation (\ref{5.59a}) shows that
$k(x)$ is the Levi-Civita connection for the tetrad field $h(x)$.
Substitution of the equations (\ref{5.58a}) into the equations
(\ref{5.59b}) leads to the familiar Einstein equations.

Turn now to the fermion matter.
Given the $L_s$-principal lift $P_\Si$ of $LX_\Si$,
let us consider the composite spinor bundle $S$ (\ref{L1}).

Set up the principal connection onthe bundle $LX_\Si$ which is given by
the local connection form
\ben &&A_\Si=(\wt A^{ab}{}_\mu dx^\mu+A^{ab}{}^c_\mu d\si^\mu_c)\otimes I_{ab},
\label{L10}\\
&&\wt A^{ab}{}_\mu=\frac12 K^\nu{}_{\la\mu}\si^\la_c (\eta^{cb}\si^a_\nu
-\eta^{ca}\si^b_\nu ),\nonumber\\
&&A^{ab}{}^c_\mu=\frac12(\eta^{cb}\si^a_\mu -\eta^{ca}\si^b_\mu),\label{M4}
\een
where $K$ is some symmetric connection on $TX$ and (\ref{M4})
correspond to the canonical left-invariant free-curvature connection on
the bundle \[GL_4\to GL_4/L.\]
The connection on the spinor bundle $S$ which is associated with
$A_\Si$ (\ref{L10}) reads
\[A_\Si=dx^\la\otimes (\dr_\la +\frac12\wt A^{ab}{}_\la
I_{ab}{}^B{}_Ay^A\dr_B) + d\si^\mu_c\otimes
(\dr^c_\mu+\frac12 A^{ab}{}^c_\mu I_{ab}{}^B{}_Ay^A\dr_B) .\]

The total configuration space of the fermion-gravitation complex is
the product
\[J^1S\op\times_{J^1\Si}J^1C_L.\]
On this configuration space, the Lagrangian density $L_{FG}$ of the
fermion-gravitation complex is the sum of the Hilbert-Einstein
Lagrangian density $L_{HE}$ (\ref{5.56}) and the modification $L_{\wt
D}$ (\ref{229})
of the Lagrangian density (\ref{5.60}) of fermion fields:
\be &&L_{\wt D}=\{\frac{i}2[ y^+_A(\g^0\g^\m)^A{}_B( y^B_\m -\frac12
(k^{ab}{}_\mu -A^{ab}{}^c_\nu(\si^\nu_{c\mu}-\G^\nu_{c\mu}))
I_{ab}{}^B{}_{C\m}y^C) -\\
&& \quad ( y^+_{A\m}-\frac12
(k^{ab}{}_\mu -A^{ab}{}^c_\nu(\si^\nu_{c\mu}-\G^\nu_{c\mu}))
I^+_{ab}{}^C{}_{A\m}y^+_C)(\g^0\g^\m)^A{}_By^B]
-my^+_A(\g^0)^A{}_By^B\}\si^{-1}\om\ee
where
\[\G^\nu_{c\mu}=frac12k^{mn}{}_\mu(\eta_{cn}\delta^d_m-
\eta_{cm}\delta^d_n) - K^\nu{}_{\la\mu}\si^\la_c,\]
\[\g^\mu=\si^\mu_a\g^a, \si=\det(\si^\mu_a).\]

The total phase space $\Pi$ of the fermion-gravitation complex is
coordinatized by
\[(x^\la,\si^\mu_c,\si^\mu_{c\nu},y^A,y^+_A,k^{ab}{}_\mu, p^{c\la}_\mu,
p^{c\nu\la}_\mu,p^\la_A, p^{A\la}_+, p_{ab}{}^{\mu\la})\]
and admits the corresponding splitting (\ref{230}). The Legendre morphism
associated with the Lagrangian density $L_{FG}$ defines the constraint
subspace of $\Pi$ given by the relations (\ref{5.61}), (\ref{5.57a}), the
conditions
\[ p_{ab}{}^{(\la\mu)}=0,\qquad p^{c\nu\mu}_\la =0\]
and the constraint (\ref{232}) which takes the form (\ref{M2}).

Hamiltonian forms associated with the Lagrangian density $L_{FG}$ are
the sum of the Hamiltonian forms $H_{HE}$ and $H_S$ (\ref{N54}) where
\begin{equation}
A^A{}_{B\mu}=\frac12 k^{ab}{}_\mu I_{ab}{}^A{}_By^B. \label{L15}
\end{equation}
The corresponding Hamilton equations for spinor fields consist with
the equations (\ref{5.62a}) and (\ref{5.62b}) where $A$ is given by the
expression (\ref{L15}). The Hamilton equations (\ref{5.58a}) -
(\ref{5.58d}) remain true. The Hamilton equations (\ref{5.58e}) and
(\ref{5.58f}) contain additional matter sources. On the constraint space
\[p^{a\mu}_\la =0\]
the modified equations (\ref{5.58f}) would come to the familiar Einstein
equations
\[G^a_\mu +T^a_\mu=0\]
where $T$ denotes the energy-momentum tensor of fermion fields,
otherwise on the modified constraint space (\ref{M2}). In the latter case,
we have
\begin{equation}
D_\la p^{c\la}_\mu=G^a_\mu +T^a_\mu \label{L16}
\end{equation}
where $D_\la$ denotes the covariant derivative with respect to the
Levi-Civita connection which acts on the indices $^c_\mu$. Substitution
of (\ref{M2}) into (\ref{L16}) leads to the modified Einstein equations
for the total system of fermion fields and gravity:
\[-\frac12 J^\la_{ab} D_\la A^{ab}{}^c_\mu=G^c_\mu +T^c_\mu\]
where $J$ is the spin current of the fermion fields.

\end{document}